# Spin Torque on Magnetic Textures Coupled to the Surface of a Three-Dimensional Topological Insulator

Ji Chen [1], Mansoor Bin Abdul Jalil[1,2], Seng Ghee Tan [1,3, a)]

[1] *Computational Nanoelectronics and Nano-device Laboratory, Electrical and Computer Engineering Department, National University of Singapore, 4 Engineering Drive 3, Singapore 117576*

[2] *Information Storage Materials Laboratory, Electrical and Computer Engineering Department, National University of Singapore, 4 Engineering Drive 3, Singapore 117576*

[3] *Data Storage Institute, A*STAR (Agency for Science, Technology and Research), DSI Building, 5 Engineering Drive 1, Singapore 117608*

## Abstract

We investigate theoretically the spin torque and magnetization dynamic in a thin ferromagnetic (FM) layer with spatially varying magnetization. The FM layer is deposited on the surface of a topological insulator (TI). In the limit of the adiabatic relaxation of electron spin along the magnetization, the interaction between the exchange interaction and the Rashba-like surface texture of a TI yields a topological gauge field. Under the gauge field and an applied current, spin torque is induced according to the direction of the current. We derived the corresponding effective anisotropy field and hence the modified Landau-Lifshitz-Gilbert equation, which describes the spin torque and the magnetization dynamic. In addition, we study the effective field for exemplary magnetic textures, such as domain wall, skyrmion, and vortex configurations. The estimated strength of the effective field is comparable to the switching fields of typical FM materials, and hence can significantly influence the dynamics of the FM layer.

Email:
TAN_Seng_Ghee@dsi.a-star.edu.sg

# Introduction

A topological insulator (TI) is a new quantum phase of condensed mater system, which has been recently attracting significant interest, both theoretically [1-4] and experimentally [5-6]. Three dimensional (3D) topological insulators, such as $Bi_{1-x}Sb_x$, $Bi_2Te_3$, and $Bi_2Se_3$, have been discovered hosting a helical metallic two dimensional electron gas (2DEG) on their surfaces. These materials act as bulk insulators but have topologically protected surface states [3-4]. Such surface states are described by the two dimensional (2D) massless Dirac equations in which the spins of the Dirac fermions are strongly coupled to their momentum. Thus, the spin orbit coupling (SOC) term is dominant in the Dirac Hamiltonian, which has a Rashba-like form. The particular topologically protected electronic surface states of the 3D TI make the system distinct from the conventional band insulator, and the 2D Dirac electron gas formed on the surface of the 3D TI is also different from the conventional electron gas in solid state systems. In the field of spintronics, TI materials constitute a promising candidate to realize novel functionality because of the spin-momentum locking property of the helical TI surface states. A number of theoretically predicted spin-related transport phenomena occurring on the 2D surfaces of TI system have already been reported in the literature, such as inverse spin-galvanic effect [7], giant spin rotation [8], giant spin battery effect [9], anomalous magnetoresistace [10], and topological charge pumping effect [11], some of which may potentially be harnessed in the next generation spintronic applications.

Spin transfer torque (STT) is a mechanism in which the angular momentum of a spin current injected into a ferromagnet is transferred to the local magnetization [12-13]. The effect opens up new possibilities to drive magnetization dynamics by electrical means, as opposed to the conventional magnetic field control. The ability to use STT to switch the magnetization is referred to as the current-induced magnetization switching (CIMS) and could remarkably improve magnetic random access memory (MRAM) technology by enabling faster magnetic switching and higher device density.

Current-induced STT effect has also been predicted in 3D TI whose surfaces are coupled to a ferromagnetic (FM) layer [14-16]. The spin-momentum locking characteristic of the 2D massless Dirac electrons on the surfaces of 3D TI suggests a strong STT effect can be achieved. The current-induced STT in 3D TI can trigger the dynamics of the magnetization of the adjacent FM layer, and even achieve magnetization reversal under certain conditions [14-15]. This kind of the current-induced STT has been investigated in a heterojuction of a 3D TI and a monodomain FM layer. The FM layer plays the role of inducing a Dirac mass, which opens a spatially uniform gap in the TI [17- 18]. The current-induced STT is thus a FM electromagnetic response through the interaction between local moments and the gapped surface Dirac electrons. However, the current-induced STT effect at the interface of TI and FM layer with spatially nonuniform magnetization (e.g. domain wall) has largely not been studied. In such spatially varying FM systems, time reversal symmetry (TRS) is broken by the inhomogeneous magnetization and a spatially dependent energy gap is opened on the surface of the TI. In this paper, we introduce a gauge field formalism [19] to study the dynamics of the spatially varying magnetization coupled to the surface state of a TI, and derive the effective field and hence the modified Landau-Lifshitz-Gilbert (LLG) equation in the presence of an applied charge current.

## Theory

We consider the spatially varying magnetization dynamics of an insulating FM layer deposited on top of a TI, as shown in Fig. 1. The low energy effective Hamiltonian for the TI surface state, including the spin-exchange interaction between the surface conduction fermions in TI and the localized spins in the FM layer, is given by,

$$\mathrm{H} = v_F(\sigma_y p_x - \sigma_x p_y) + J\boldsymbol{\sigma}\cdot\boldsymbol{M}(\boldsymbol{r}),\qquad(1)$$

In the above Hamiltonian, the first term is the 2D massless Dirac-Rashba Hamiltonian, where $v_F$ is the Fermi velocity, $\boldsymbol{\sigma}=(\sigma_x,\sigma_y,\sigma_z)$ is the vector of Pauli matrices that act on the spin degree of freedom, and $p_i = -i\hbar \nabla_i$ ($i = x, y$) is the momentum operator acting on the conduction electron. The second term is the exchange interaction between the local spins and the surface Dirac fermions, where $J$ is the exchange coupling integral and $\boldsymbol{M}(\boldsymbol{r})$ is the local magnetization due to the localized d-orbitals in the magnetic materials and is spatially dependent. At the Dirac point where the momentum of the surface fermion is zero, the ground state will be TRS broken by a finite z-component of the magnetization $M_z(\boldsymbol{r})$ and an energy gap will be opened [20]. We now begin to investigate the FM response through the interaction between the local spins and the gapped surface Dirac electrons of the TI in the presence of the applied charge current.

In order to look at the spatially varying system in a convenient way, it is necessary to perform a local gauge transformation, which locks the reference spin quantization axis to the local magnetization $\boldsymbol{M}(\boldsymbol{r})$ at each point in $\boldsymbol{r}$-space (Fig. 2). In the laboratory frame, as an electron moves from one point to another, it can be seen that the reference spin axis rotates in $\boldsymbol{r}$-space and is always parallel to the direction of the local magnetization axis $\boldsymbol{n} = \boldsymbol{M}(\boldsymbol{r})/\|\boldsymbol{M}(\boldsymbol{r})\|$. In the rotated reference frame, the reference spin quantization axis is fixed to $\boldsymbol{z}$ along $\boldsymbol{n}$ at every point in $\boldsymbol{r}$-space. Thus, after the gauge transformation, the problem is effectively equivalent to a spatially uniform system. We employ a unitary rotation matrix $U = U(\boldsymbol{r})$, which satisfies $U(\boldsymbol{\sigma} \cdot \boldsymbol{n})U^\dagger = \sigma_z^r$ at every point in $\boldsymbol{r}$-space, where $\sigma_z^r = \sigma_z = \begin{pmatrix} 1 & 0 \\ 0 & -1 \end{pmatrix}$ is the new z-Pauli matrix in the reference frame and the superscript $r$ denotes the new reference frame. The rotation unitary matrix is given by

$$U = \boldsymbol{m} \cdot \boldsymbol{\sigma}, \qquad (2)$$

where $\boldsymbol{m} = \left(\sin\dfrac{\theta}{2}\cos\phi,\ \sin\dfrac{\theta}{2}\sin\phi,\ \cos\dfrac{\theta}{2}\right)$ and $(\theta(\boldsymbol{r}), \phi(\boldsymbol{r}))$ are the polar and azimuthal orientations of the local magnetization in spherical coordinates that parameterizes the vector $\mathbf{n} = (\sin\theta\cos\phi, \sin\theta\sin\phi, \cos\theta)$, and are functions of position $\mathbf{r}$. The effective Hamiltonian after gauge transformation becomes

$$\begin{aligned}
H' = UHU^\dagger &= Uv_F(\sigma_y p_x - \sigma_x p_y)U^\dagger + UJ\boldsymbol{\sigma}\cdot\mathbf{M}(\mathbf{r})U^\dagger \\
&= v_F(U\sigma_y U^\dagger p_x + U\sigma_y(p_x U^\dagger)) - v_F(U\sigma_x U^\dagger p_y + U\sigma_x(p_y U^\dagger)) \\
&\quad + J|M(\mathbf{r})|\sigma_z^r,
\end{aligned} \quad (3)$$

where $|M(\mathbf{r})| = \sqrt{M_x^2(\mathbf{r}) + M_y^2(\mathbf{r}) + M_z^2(\mathbf{r})}$. In the rotated reference frame, the reference axis is rotated to $(0\ 0\ 1)$, and the exchange interaction term $J\boldsymbol{\sigma}\cdot\mathbf{M}(\mathbf{r})$ is diagonalized.

The transformed Hamiltonian includes the contributions from the kinetic energy, interaction energy, as well as the exchange coupling energy. However, we cannot simply regard the first term, $v_F(U\sigma_y U^\dagger p_x - U\sigma_x U^\dagger p_y)$, as the kinetic energy and the second term $v_F(U\sigma_y p_x U^\dagger - U\sigma_x p_y U^\dagger)$, as the interaction energy, since after examining the Hermiticity of the above terms, we found that they are all non-Hermitian. In quantum mechanics, the mechanical operators that are associated with the measurable properties must be Hermitian to ensure their eigenvalues are real (not complex), since the measurement outcome of any experiment must be a real number. Non-Hermitian operators would not correspond to observable properties, such as the operators $v_F(U\sigma_y U^\dagger p_x - U\sigma_x U^\dagger p_y)$ and $v_F(U\sigma_y(p_x U^\dagger) - U\sigma_x(p_y U^\dagger))$ in the transformed Hamiltonian (3). The possible reason for the operators being non-Hermitian is that the terms $v_F(U\sigma_y U^\dagger p_x - U\sigma_x U^\dagger p_y)$ and $v_F(U\sigma_y(p_x U^\dagger) - U\sigma_x(p_y U^\dagger))$ include part of the kinetic energy, as well as part of the interaction energy. Thus, the sum of the above two operators is Hermitian, as it represents both the kinetic energy and interaction energy, which is a measurable quantity.

To separate out the kinetic and interaction energy terms such that both are represented by Hermitian operators, we first consider the velocity operator for the rotated system, which is obtained from Hamilton's equation of motion as follows [21]

$$\boldsymbol{v}(\boldsymbol{r}) = \frac{\partial \mathrm{H}'}{\partial \boldsymbol{p}} = v_F(U\sigma_y U^\dagger, -U\sigma_x U^\dagger). \tag{4}$$

The velocity operator is spatially dependent since the rotation unitary matrix $U$ is varying in real space. We then introduce the spatially varying velocity in a symmetric way into the kinetic energy operator, that is,

$$\boldsymbol{v}(\boldsymbol{r})\boldsymbol{p} \Rightarrow \frac{\boldsymbol{v}(\boldsymbol{r})\boldsymbol{p} + \boldsymbol{p}\boldsymbol{v}(\boldsymbol{r})}{2}, \tag{5}$$

such that $\boldsymbol{v}(\boldsymbol{r})\boldsymbol{p}$ is a Hermitian operator, where $\boldsymbol{p} = (p_x, p_y)$. The transformed Hamiltonian can thus be reformulated to read

$$\begin{aligned}\mathrm{H}' \equiv v_F &\left\{\frac{1}{2}U\sigma_y U^+[p_x + U(p_x U^+)] + \frac{1}{2}[p_x + U(p_x U^+)]U\sigma_y U^+\right\} \\ -v_F &\left\{\frac{1}{2}U\sigma_x U^+[p_y + U(p_y U^+)] + \frac{1}{2}[p_y + U(p_y U^+)]U\sigma_x U^+\right\} \\ &+ J\sigma_z^r |M(\boldsymbol{r})|,\end{aligned} \tag{6}$$

where the terms $v_F/2\left[U\sigma_y U^+ p_x + p_x\left(U\sigma_y U^+\right)\right] - v_F/2\left[U\sigma_x U^+ p_y + p_y\left(U\sigma_x U^+\right)\right]$ contribute to the kinetic energy, and arise from the unitary transformation and the symmetrization of the kinetic energy operator in Eq. (5), while the terms $v_F/2[U\sigma_y U^+\left(U\left(p_x U^+\right)\right) + U\left(p_x U^+\right)\left(U\sigma_y U^+\right)] - v_F/2[U\sigma_x U^+\left(U\left(p_y U^+\right)\right) + U(p_y U^+)\left(U\sigma_x U^+\right)]$ contribute to the interaction energy, which arises due to the unitary transformation. The interaction energy is generated by the coupling of the incident electrons (the injected current) to the local magnetization in the presence of the Rashba-like SOC on the TI surface.

Now we assume that the local magnetization is smoothly varying in the system and the motion of the injected electron is sufficiently slow, in order to allow its spin orientation to relax to the magnetization direction, i.e., to align parallel (antiparallel) to the localized spin. Thus, we are considering the so-called Abelian (adiabatic) approximation, in which the interband transitions between the two spin eigenstates of the carriers (i.e., spin up and spin down eigenstates) are forbidden and the carriers have to remain in their spin eigenstates in the quantum system [22]. Mathematically, this is identical to setting the off-block-diagonal matrix elements of the transformed Hamiltonian that connect the spin up and spin down bands to zero, and to retain only the diagonal components of the transformed Hamiltonian. After Abelian (adiabatic) approximation, we arrive at the simplified Hamiltonian:

$$H' \approx \frac{v_F}{2} \{(\sin\theta\sin\phi p_x + p_x \sin\theta\sin\phi)\sigma_z^r - (\sin\theta\cos\phi p_y + p_y \sin\theta\cos\phi)\sigma_z^r\}$$

$$+ (ev_F \sin\theta\sin\phi A_x - ev_F \sin\theta\cos\phi A_y) + J|M(r)|\sigma_z^r, \qquad (7)$$

where the first term in the Hamiltonian contribute to the kinetic energy and the second term contribute to the interaction energy. The velocity operator for the system is also obtained from Hamilton's equation of motion

$$v(r) = \frac{\partial H'}{\partial p} = (v_F \sin\theta\sin\phi\sigma_z^r, -v_F \sin\theta\cos\phi\sigma_z^r), \qquad (8)$$

and $A = (A_x, A_y) = \left(\hbar/2e(\partial_x\phi - \cot\phi/\sin\theta\partial_x\theta)\sigma_z^r, \hbar/2e(\partial_y\phi + \tan\phi/\sin\theta\partial_y\theta)\sigma_z^r\right)$ is the gauge field (electromagnetic vector potential) that is induced by the combined effect of the SOC (the spin-momentum locking) on the TI surface and coupling to the local magnetization of the FM layer. The approximated Hamiltonian can be explained as a result of the competition between the exchange interaction and the Rashba-like surface texture, which yields a finite component of $\sigma^r$ in the z direction in the reference frame. It gives a simple physical picture of the behaviour of the surface Dirac electrons of TI when they interact with a spatially uniform local magnetization in the rotated reference frame. The momentum of the surface Dirac electrons is thus modified $p \to p + A$. Notably, the gauge field $A$ is a $2\times 2$ matrix and can be described by a $U(1)\times U(1)$ Abelian gauge field (i.e., whose components commute), which describes the motion of the conduction electrons in the given field $n(r)$. It can be seen from the gauge field $A = A_{11}\sigma_z^r$ (where the subscript 11 denotes the top left diagonal element of the gauge field matrix), i.e., in the reference frame, the up- and down-spin Dirac fermions on the TI surface experience effective gauge fields with the same strength but of opposite signs. The gauge field leads to an effective locally varying magnetic field $\bm{B}(\bm{n}(\bm{r})) = \nabla \times \bm{A}$ in the adiabatic limit, which is pointing in the z-direction on the x-y plane of the medium. Such a gauge field will affect the electron transport properties in several significant ways, such as spin separation and spin electron motive force [23], and is related to the spin Hall [22] and anomalous Hall effects [24].

The applied current $\bm{j}$ interacting with the electromagnetic vector potential $\bm{A}$ is a source of the current-driven spin torque in the 3D TI system. The electromagnetic interaction energy between the fermion field (of the injected electron) and the gauge field (vector potential) due to the combined presence of the SOC on the TI surface and coupling to the local magnetization, is given by

$$E_{\text{int}} = \int \psi'^{\dagger} e v_F (\sin\theta \sin\phi A_x - \sin\theta \cos\phi A_y) \psi' d^2 x, \qquad (9)$$

where the current operator of the TI system in the reference frame is defined as $\bm{j} = e v_F \psi'^{\dagger} (\sin\theta \sin\phi \sigma_z^r, -\sin\theta \cos\phi \sigma_z^r) \psi'$ and $\psi' = U\psi$ is the wavefunction of the system in the reference frame. In the adiabatic limit (strong exchange coupling $J \to \infty$), we assume a fully spin-polarized charge current (along $\bm{M}$) propagating in the region of the spatially varying $\bm{M}$. All of the incident electron spins are in the spin-up state $\psi'_{\uparrow}$ and experience a gauge field $A_{\uparrow} = \langle \psi'_{\uparrow} | A | \psi'_{\uparrow} \rangle = A_{11}$, where $A_{11} = (A_{11}^x, A_{11}^y) = (\hbar/2e(\partial_x \phi - \cot\phi/\sin\theta \, \partial_x \theta), \hbar/2e(\partial_y \phi + \tan\phi/\sin\theta \, \partial_y \theta))$ is the top-left diagonal element of the matrix $A$. The interaction energy between the current and the local magnetization is then given by

$$E_{\text{int}} = \int \psi'^{\dagger}_{\uparrow} e v_F \left( \sin\theta \sin\phi A_{11}^x - \sin\theta \cos\phi A_{11}^y \right) \psi'_{\uparrow} d^2 x \equiv \int \bm{j}_{\uparrow} \Box A_{11} \, d^2 x. \qquad (10)$$

Physically, $A_{11}$ is a TI SOC induced magnetic vector potential in the reference frame that affects the energy and the dynamics of spin-up electrons only, while $\bm{j}_{\uparrow} = (j_x^{\uparrow}, j_y^{\uparrow}) = \psi'^{\dagger}_{\uparrow} e v_F (\sin\theta \sin\phi, -\cos\theta \sin\phi) \psi'_{\uparrow}$ is the electric charge current density under extremely large exchange coupling strength $J$, such that it is fully polarized along the local magnetization direction. In micromagnetic analysis, the local magnetization will adjust its orientation to achieve the minimum energy at equilibrium. An effective anisotropy magnetic field due to the coexistence of the applied charge current, the SOC in TI and the local magnetization can thus be obtained by taking the energy gradient with respect to the local magnetization $M(r)$, i.e., $H^{eff} = -\delta E_{int}/\mu_0 \delta M(r)$ [25]. Note that since the interaction energy is expressed as a functional, the effective field $H^{eff}$ is given by a functional derivative, which is a generalization of the directional derivative. By considering the energy functional in Eq. (11), the current-driven effective magnetic field is then explicitly given by

$$H^{eff} = -\frac{\delta E_{int}}{\mu_0 \delta M(r)} = \frac{1}{\mu_0 M}\frac{\hbar}{2e}\left(\frac{j_x}{n_y^2},\frac{j_y}{n_x^2},0\right)\cdot(\nabla n \times n), \quad (11)$$

where $n = M/M = (n_x, n_y, n_z)$ is the unit vector of the local magnetization, $M$ is the saturation magnetization and $\mu_0 = 4\pi \times 10^{-7} TmA^{-1}$ is the vacuum permeability. This current-induced effect contains the information from both the SOC in TI and spin relaxation to the spatially varying magnetization. The term $(\nabla n \times n)$ is a gauge transform-induced effect while the term $\left(\frac{1}{n_y^2},\frac{1}{n_x^2},0\right)$ originates from the Rashba-type SOC in TI. The effective magnetic field $H^{eff}$ can be regarded as an externally applied magnetic field, which will affect the dynamics of the magnetization of the FM layer.

The magnetization motion is governed by the LLG equation. We can write the dynamic equation of the magnetization by including the above current-induced effective magnetic field $H^{eff}$:

$$\frac{dM}{dt} = -\gamma M \times (H + H^{eff}) + \alpha M \times \frac{dM}{dt}, \quad (12)$$

where $\gamma$ is the gyromagnetic ratio with units of $mA^{-1}s^{-1}$, $\alpha$ is the Gilbert damping parameter and $H = H^{exch} + H^{mag} + H^{ani}$ is the effective field due to the contributions from the exchange energy, the magnetostatic energy, and the anisotropy energy. In the above Eq. (12), the first term is the precession term, in the low damping limit, the local magnetization will precess about a total effective field $\boldsymbol{H} + \boldsymbol{H}^{\text{eff}}$ and the second term is the usual damping term in the LLG equation. The current-induced spin torque is defined as $\boldsymbol{T} = -\gamma \boldsymbol{M} \times \boldsymbol{H}^{\text{eff}}$ with $\boldsymbol{H}^{\text{eff}} = \frac{1}{\mu_0 M} \frac{\hbar}{2e} \left( \frac{j_x}{n_y^2}, \frac{j_y}{n_x^2}, 0 \right) (\boldsymbol{\nabla} \boldsymbol{n} \times \boldsymbol{n})$, which represents the current-driven field acting on the local magnetization $\boldsymbol{M}$.

## Results and Discussions

We consider the specific configuration shown in Fig. 1, where the current is propagating along $x$-direction. Thus, all the spatial derivatives reduce to $\boldsymbol{\nabla} \to \partial_x$. The current induced effective magnetic field is thus given by

$$\boldsymbol{H}^{\text{eff}} = \frac{\hbar}{2\mu_0 M} \frac{j_x}{e} \frac{1}{n_y^2} (\partial_x \boldsymbol{n} \times \boldsymbol{n}). \tag{13}$$

Let us assume that the FM layer possesses a one-dimensional domain wall (DW) configuration, which is expressed in polar angles by

$$\boldsymbol{n}(\boldsymbol{r}) = (\sin\theta\cos\phi, \sin\theta\sin\phi, \cos\theta)$$

$$\theta(\boldsymbol{r}) = 2\tan^{-1} e^{-(x-x_{\text{dw}})/\lambda_{\text{dw}}}, \quad \phi(\boldsymbol{r}) = \phi_{\text{dw}}, \tag{14}$$

where $\boldsymbol{r} = (x, y)$ is the position parameter, $\lambda_{\text{dw}}$ is the DW width, $x_{\text{dw}}$ and $\phi_{\text{dw}}$ are the central position and the azimuthal angle of DW, respectively. The DW is moving along the $y$ axis, and the magnetization direction $\boldsymbol{n}$ is pointing in the $\pm z$ directions at $x = \pm\infty$. When $\phi(\boldsymbol{r}) = \pm\pi/2$, we have a Bloch wall, and the magnetization direction $\boldsymbol{n}$ can be expressed as $\boldsymbol{n} = (0, \pm\sin\theta(\boldsymbol{r}), \cos\theta(\boldsymbol{r}))$. The current-induced effective magnetic field will then point in the $x$ direction, i.e.,

$$H_{\text{eff}}^x = \frac{1}{\mu_0 M} \frac{\hbar}{2e} \frac{j_x}{n_y^2} (n_z \partial_x n_y - n_y \partial_x n_z), \tag{15}$$

and this effective field will exert a torque on the magnetization $M_y$ and $M_z$. The current-induced magnetic fields are functions of electron's position $x$ in the DW. From Eqs. (14) and (15), the effective field is explicitly given by

$$H^x_{\text{eff}} = \mp \frac{j_x \hbar}{\mu_0 e M \lambda_{\text{dw}}} \frac{1}{\sin^2[2\tan^{-1} e^{-(x-x_{\text{dw}})/\lambda_{\text{dw}}}]} \frac{e^{-(x-x_{\text{dw}})/\lambda_{\text{dw}}}}{1+e^{-2(x-x_{\text{dw}})/\lambda_{\text{dw}}}}. \quad (16)$$

It can be seen that the effective field has the order of $H^x_{\text{eff}} \Box \frac{j_x \hbar}{\mu_0 e M \lambda_{\text{dw}}}$. There is an inverse relationship with the DW width, i.e., the current-induced effective magnetic field strength increases with decreasing $\lambda_{\text{dw}}$. Assuming a typical DW width of $\lambda_{\text{dw}} = 24$ nm, and a current density of $j_x = 10^8 \, Acm^{-2}$, and taking the DW position $x_{\text{dw}} = 0$, one arrive at an estimated effective field of $H^x_{\text{eff}} \Box 5.45 \times 10^5 \, Am^{-1}$ for the Co material with the saturation magnetization $M_{Co} = 1.09 \times 10^6 \, Am^{-1}$. For Fe and Ni materials, the estimated effective field is $H^x_{\text{eff}} \Box 3.47 \times 10^5 \, Am^{-1}$ for $M_{Fe} = 1.71 \times 10^6 \, Am^{-1}$ and $H^x_{\text{eff}} \Box 1.2 \times 10^6 \, Am^{-1}$ for $M_{Ni} = 4.9 \times 10^5 \, Am^{-1}$, respectively. The current-induced fields are sufficient to trigger the dynamics of the above FM materials like Co, Fe and Ni, since tits strength is comparable to the switching field for these materials, viz. $H^C_{Co} = 7429 \, Oe \, (5.91 \times 10^5 Am^{-1})$, $H^C_{Fe} = 565 Oe \, (4.5 \times 10^4 Am^{-1})$ and $H^C_{Ni} = 233 \, Oe \, (1.85 \times 10^4 Am^{-1})$. Next, we consider the Neel wall configuration, i.e., when $\phi(\boldsymbol{r}) = 0$ or $\pi$, such that the magnetization vector in spherical coordinates is $\boldsymbol{n} = (\pm\sin\theta(\boldsymbol{r}), 0, \cos\theta(\boldsymbol{r}))$. Current in the $x$-direction will induce an effective field in the $y$-direction, given by

$$H^y_{\text{eff}} = \frac{1}{\mu_0 M} \frac{\hbar}{2e} \frac{j_x}{n_y^2} (-n_z \partial_x n_x + n_x \partial_x n_z). \quad (17)$$

This effective magnetic field will exert a torque on the magnetization components $M_x$ and $M_z$. We assume there is a small fluctuation of the local magnetization for this case, i.e., $n_y = \varepsilon$, where $\varepsilon$ is a some small number. This leads to a large $y$ component of the current-induced effective field $H_{\text{eff}}^{y} = \frac{1}{\mu_0 M} \frac{\hbar}{2e} \frac{j_x}{\varepsilon^2}(-n_z \partial_x n_x + n_x \partial_x n_z)$, regardless of the FM material.

We next study the magnetic dynamics of the 2D magnetic configuration in a thin FM on top of the TI. For magnetic configuration with a radially symmetric $z$-component, e.g., skyrmion or vortex configuration, the magnetic orientation may be expressed as $n(r) = (\sin\theta \cos\phi, \sin\theta \sin\phi, \cos\theta)$, where $\theta=\theta(\rho)$, $\phi=\phi(\omega)=W\omega$, the integer $W$ representing the winding number. Note that $(\theta,\phi)$ are the coordinates that describe the magnetization orientation on the Bloch sphere in spin space, while $(r,\omega)$ are the cylindrical coordinates in real space satisfying $r=(x,y,z)=(\rho\cos\omega, \rho\sin\omega, z)$. In the presence of an applied current in the $x$-direction which interacts with a generalized 2D vortex/skyrmion configuration, the resulting effective magnetic field has the following $x$, $y$ and z components:

$$H_{\text{eff}}^{x} = \frac{j_x \hbar}{2e\mu_0 M}\left[\frac{\cos\omega}{\sin^2\theta(r)\sin(W\omega)} - \frac{W\cos\theta(r)\cos(W\omega)}{\sin\theta(r)\sin^2(W\omega)}\frac{\sin(\omega)}{\rho}\right],$$

$$H_{\text{eff}}^{y} = \frac{j_x \hbar}{2e\mu_0 M}\left[\frac{-\cos W\omega \cos\omega}{\sin^2\theta(r)\sin^2 W\omega} - \frac{W\cos\theta(r)}{\sin\theta(r)\sin W\omega}\frac{\sin\omega}{\rho}\right],$$

$$H_{\text{eff}}^{z} = \frac{j_x \hbar}{2e\mu_0 M}\frac{1}{\sin^2(W\omega)} W \frac{\sin\omega}{\rho}. \tag{18}$$

This effective field induces a torque which affects the dynamics of all three components of the local magnetization $M_x$, $M_y$ and $M_z$. For the specific skyrmion field expressed by $\cos\theta(r) = \frac{\rho^2 - a^2}{\rho^2 + a^2}$ and $\phi(\omega) = W\omega$, where $a$ is the domain size and $W$ is the winding number, the current driven field is obtained by

$$H_{\text{eff}}^{x} = \frac{j_x \hbar}{2e\mu_0 M}\left[\frac{\cos\omega(\rho^2+a^2)^2}{4\rho^2 a^2 \sin(W\omega)} - \frac{W(\rho^2-a^2)\cos(W\omega)\sin\omega}{2\rho^2 a \sin^2(W\omega)}\right],$$

$$H_{\text{eff}}^{y} = \frac{j_x \hbar}{2e\mu_0 M}\left[\frac{-\cos W\omega \cos\omega(\rho^2+a^2)^2}{4\rho^2 a^2 \sin^2 W\omega} - \frac{W(\rho^2-a^2)\sin\omega}{2\rho^2 a \sin W\omega}\right],$$

$$H^z_{\text{eff}} = \frac{j_x \hbar}{2e\mu_0 M} \frac{1}{\sin^2(W\omega)} W \frac{\sin\omega}{\rho}. \quad (19)$$

Assuming a winding number $W=1$ and the skyrmion configuration with boundary at $\rho = 10$ nm, the strength of the effective field strength is of the order of $\boldsymbol{H}_{\text{eff}} \Box \frac{j_x \hbar}{2e\mu_0 M \rho} \approx 1 \times 10^5 \text{Am}^{-1}$ for $M_{Ni} = 4.9 \times 10^5 \text{ Am}^{-1}$, under the applied current density $j_x = 10^7 A/cm^{-2}$. For the vortex configuration, represented by $\cos\theta(\rho) = \frac{Pa^2}{\rho^2 + a^2}$ and $\phi(\omega) = W\omega + (-1)^C \times \frac{\pi}{2}$, where $P = \pm 1$ is the polarity, $C = \pm 1$ is the chirality and $a$ is the domain size, the current driven field is given by

$$H^x_{\text{eff}} = \frac{j_x \hbar}{2e\mu_0 M} \left[ \frac{\cos\omega(\rho^2 + a^2)^2}{((\rho^2+a^2)^2 - P^2 a^2)\sin(W\omega)} - \frac{WPa\cos(W\omega)\sin(\omega)}{\rho\sqrt{(\rho^2+a^2)^2 - P^2 a^2}\sin^2(W\omega)} \right],$$

$$H^y_{\text{eff}} = \frac{j_x \hbar}{2e\mu_0 M} \left[ \frac{-\cos W\omega \cos\omega(\rho^2 + a^2)^2}{((\rho^2+a^2)^2 - P^2 a^2)\sin^2 W\omega} - \frac{WPa\sin\omega}{\rho\sqrt{(\rho^2+a^2)^2 - P^2 a^2}\sin W\omega} \right],$$

$$H^z_{\text{eff}} = \frac{j_x \hbar}{2e\mu_0 M} \frac{1}{\sin^2(W\omega)} W \frac{\sin\omega}{\rho}. \quad (20)$$

The vortex configuration will induce an effective field of comparable strength to that induced by the skyrmion configuration, i.e., $\boldsymbol{H}_{\text{eff}} \Box \frac{j_x \hbar}{2e\mu_0 M \rho} \approx 1 \times 10^5 \text{ Am}^{-1}$ for $M_{Ni} = 4.9 \times 10^5 \text{ Am}^{-1}$ assuming a current density $j_x = 10^7 A/cm^{-2}$, and for simplicity, winding number $W = 1$, the polarity $P = 1$ and $C = 1$. For both the skymion and vortex configurations, it can be seen that the strength of the current induced effective field increases as the radius of the configuration decreases. Therefore, for the atomic scale skyrmions (with radius of 1 nm) in monolayer Fe on Ir (111), which was discovered recently [26], the current driven effective field is potentially even larger.

# Conclusion

We have presented a gauge formalism to describe the current-induced dynamics of a ferromagnetic film with spatially varying magnetization, which is deposited on top of a three-dimensional topological insulator. An applied current is injected onto the surface of the 3D TI and the combination of *s-d* coupling and Rashba-like SOC on the TI surface induces an effective magnetic field under adiabatic condition. The effective field modifies the LLG equation, which governs the magnetization dynamics of the FM film. We investigate the current-induced field for some exemplary magnetization one- (Neel and Bloch domain walls) and two-dimensional (skyrmion and vortex) configurations in the FM film. We obtain an estimate of the field strength which is comparable to the switching fields of typical FM materials.

# Acknowledgement

**Figure Captions**

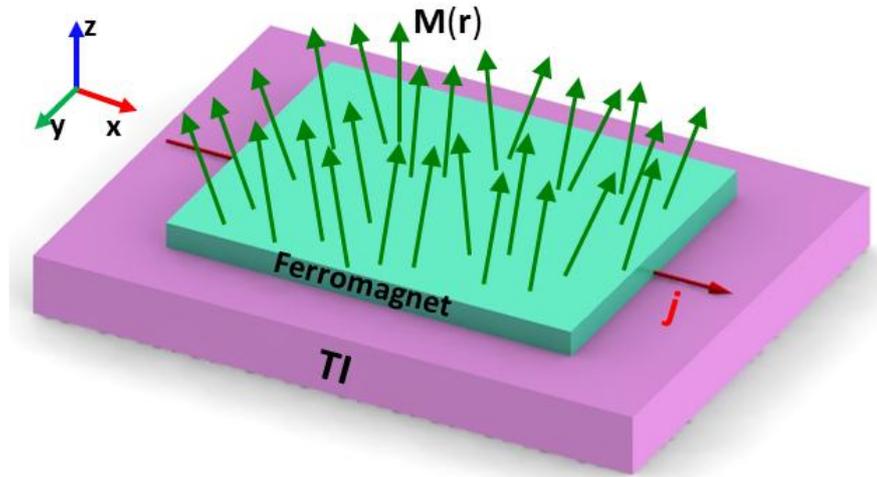

Fig.1 (Color online) Schematic of a FM layer with the spatially varying local magnetization deposited on top of a topological insulator. An applied current is propagating on the surface of the TI. The existence of the FM layer breaks the TRS of the TI surface and opens up a spatially dependent gap on the surface of the TI.

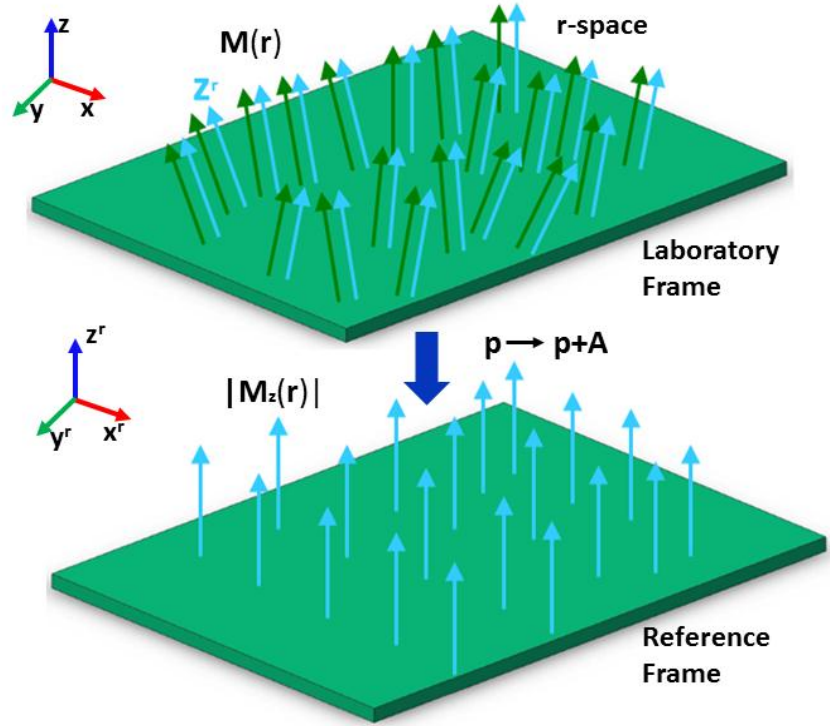

Fig.2 (Color online) Schematic of the local gauge transformation. The green arrows denote the local magnetization axis in laboratory frame and the blue arrows denote the reference spin quantization axis $z^r$ in reference frame. A spatially varying FM system (top) can be effectively transformed to a uniform system (bottom) by the local gauge transformation. This transformation results in a modification of the carrier's momentum from $p$ to $p+A$, where $A$ is a gauge field (electromagnetic vector potential) due to the interaction between the SOC in TI and local magnetization.